\documentstyle[12pt,epsf,epsfig]{article}

\def\bm#1        {{\bf #1}}
\def\to          {\rightarrow}

\def\GeVc2       {{\rm GeV}/c^2}

\def\lapproxeq   {\lower .7ex\hbox{$\;\stackrel{\textstyle
                  <}{\sim}\;$}}
\def\gapproxeq   {\lower .7ex\hbox{$\;\stackrel{\textstyle
                  >}{\sim}\;$}}
\def\beq         {\begin{equation}}
\def\eeq         {\end{equation}}

\def\pom     {\mbox{{\scriptsize P}\hspace{-0.16cm}{\scriptsize I}}}
 
\def\bigpom  {\mbox{ P\hspace{-0.22cm}I}}

\def\calP    {{\cal P}}

\def\xbj     {x_{\rm Bj}}
\def\xp      {x_{\cal P}}
\def\xpom    {x^{\rm\pom}}

\begin{document}

\begin{titlepage}

\begin{flushright}  
ETH--TH/97--6      \\
DTP/97/08          \\
hep-ph/9702286     \\
January 1997       \\
\end{flushright}   

\vskip   1.2cm   

\begin{center}
{\Large\bf  Diffractive Higgs Production at the LHC} 
\vskip 2.cm 

{\large M.~Heyssler}
\vskip .3cm
{\it Department of  Physics, University of Durham \\
Durham DH1 3LE, England }\\
\vskip .5cm
{\large Z.~Kunszt}
\vskip .3cm
{\it Institute of Theoretical Physics, ETH\\
 CH--8093 Zurich, Switzerland}\\
\vskip .3cm
and
\vskip .3cm
{\large W.J.~Stirling}
\vskip .3cm
{\it Departments of Mathematical Sciences and Physics, 
University of Durham \\
Durham DH1 3LE, England }\\

\vskip 1cm

\end{center}

\begin{abstract}
We use  diffractive  parton  distributions  obtained  from fits to the
diffractive  structure  function  measured  at HERA to  predict  cross
sections  for single  diffractive  Higgs  production  at the LHC.  The
dominant background  processes are also considered.  Although some 5\%
-- 15\% of Higgs events are predicted to be diffractive in this model,
the ratio of signal to background is not significantly improved.
\end{abstract}

\vfill 

\end{titlepage}

\newpage

The fact that a  significant  fraction  of deep  inelastic  scattering
events seen at the HERA $ep$  collider have a  diffractive  (`rapidity
gap')  structure  has  led  to  suggestions  that  `hard   diffractive
scattering'  may be a relatively  common  occurrence  at  high--energy
lepton--hadron  and  hadron--hadron  colliders and, furthermore,  that
such  diffractive  topologies may help to enhance  certain new physics
signals over  backgrounds.  In this letter we wish to explore  further
the idea \cite{SHAFER,GRAUDENZ} that single diffractive production may
be a useful  additional  tool for  identifying  the Higgs boson at the
LHC.

An  important   property  of  the  HERA  diffractive   deep  inelastic
scattering    events~\cite{gaporig}    is   the    approximate    {\it
factorization} of the structure function $F_2^D$ (integrated over $t$)
into a  function  of  $\xp$  times  a  function  of $\xi =  \xbj/\xp$:
$F_2^{D} \sim \xp^{-n}  {\cal  F}(\xi,Q^2)$.  This property,  together
with the  observed  rapidity--gap  topology  of the  events,  strongly
suggests  that  the  deep  inelastic  scattering  takes  place  off  a
slow--moving colourless target $\calP$ `emitted' by the proton, $p \to
\calP p$, and with a  fraction  $\xp \ll 1$ of its  momentum.  If this
emission is described by a universal flux function  $f_{\calP}(\xp, t)
d\xp dt $, then the diffractive structure function $F_2^D$ is simply a
product of this and the structure  function of the colourless  object,
$F_2^\calP(\xi,Q^2)$.  Since the scattering  evidently takes place off
point--like  charged  objects,  we may write the  latter as a sum over
quark--parton distributions, i.e.  $ F_2^{\calP}(\xi,Q^2) = \xi \sum_q
e_q^2  q_{\calP}(\xi,Q^2)$, in leading order.  In this way we obtain a
model for the diffractive parton distributions:
\beq
{d  f_{q/p}(x,\mu^2;  \xp , t) \over d \xp d t} = f_{\calP}(\xp,  t)\;
f_{q/\calP}(\xi=x/\xp,\mu^2)\; .  
\label{eq:fqp}
\eeq
If one assumes  further that the  colour--neutral  target is the Regge
pomeron, then the emission factor $f_\calP$ is already known from soft
hadronic    physics    (for    a    review    see    Ref.~\cite{PVL}):
$f_{\pom}(\xpom,t) = F_{\pom}(t) (\xpom)^{2\alpha_{\pom}(t)-1}$, which
gives   a   factorized    structure    function    with   $n   \approx
2\alpha_{\pom}(0)-1  \approx 1.16$.  This model is based on the notion
of `parton constituents in the pomeron' first proposed by Ingelman and
Schlein~\cite{IS}  and supported by data from UA8~\cite{UA8}.  In such
a  model, a modest  amount  of  factorization  breaking,  such as that
observed    in   the    more    recent    H1   and    ZEUS    analyses
\cite{H1thisyear,ZEUSthisyear},  could be  accommodated  by invoking a
sum over  Regge  trajectories,  each with a  different  intercept  and
structure  function, see for example  Ref.~\protect\cite{GBK}.  In the
present study we assume, for simplicity, pomeron exchange only and use
the  parametrization  of Ref.~\cite{DL} for $  \alpha_{\pom}(t)  $ and
$F_{\pom}(t)$.  The $\xpom$  dependence of the  diffractive  structure
function  predicted  by this type of `soft  pomeron'  model is roughly
consistent   within   errors   with   the   H1~\cite{H1thisyear}   and
ZEUS~\cite{ZEUSthisyear}  data, although there is some indication that
a somewhat steeper $\xpom$ dependence is preferred.

Although the above picture of deep inelastic  scattering  taking place
off  hard  parton  constituents  in a  `soft'  pomeron  gives  a  good
description  of the  HERA  data,  the  generalization  to  other  hard
scattering  processes,  and in  particular  the concept of  `universal
pomeron  structure' is on a much less firmer theoretical  footing, see
for example Ref.~\cite{THEORY}.  One of the cleanest processes to test
this  hypothesis  would  appear to be  diffractive  $W^\pm$  and $Z^0$
production at the Tevatron, i.e.  $p \bar p$ collisions at $\sqrt{s} =
1.8$~TeV         \cite{BRUNI,Stirling96}.         According         to
Ref.~\cite{Stirling96}, some 7\% of $W$ events should exhibit a single
diffractive  structure,  that is with a  rapidity  gap in  either  the
forward or backward hemispheres.  The experimental situation is as yet
unclear,  see  for  example   Ref.~\cite{GOULIANOS}.  Using  the  same
factorization  hypothesis,  diffractive  heavy  flavour  production at
hadron colliders was studied in Ref.~\cite{MATT}.

In the present study we assume that the  universal  pomeron  structure
picture is valid, and use the quantitative  information on diffractive
parton   distributions,   extracted  from  HERA  $F_2^D$  data  as  in
Ref.~\cite{Stirling96},  to predict (single)  diffractive  Higgs cross
sections at LHC, that is Higgs production with a large rapidity gap in
one  hemisphere.\footnote{The  cross  sections for double  diffractive
production,  with two  rapidity  gaps, are readily  estimated  in this
approach by combining two sets of  diffractive  parton  distributions.
Numerically, these are found to be much smaller.  For other approaches
to double  diffractive  Higgs production see  Ref.~\protect\cite{DD}.}
The  process is depicted in Fig.~1.  This model of  diffractive  Higgs
production was first studied in  Ref.~\cite{SHAFER}.  Recently, it has
been  suggested~\cite{GRAUDENZ}  that  triggering  on single or double
diffractive  events may provide a cleaner  environment for discovering
Higgs  bosons  produced  via $gg\to H$.  The  argument  is that gluons
should be more copious in the pomeron, thus enhancing the Higgs signal
relative to the  background.  However when assessing the usefulness of
the single diffractive cross section in enhancing the Higgs signal, it
is equally important to consider the corresponding  single diffractive
{\it  background}  processes.  Naively, one might argue that since the
important backgrounds  originate in quark--antiquark  annihilation ($q
\bar q \to  \gamma\gamma,  ZZ$) the  gluon--rich  pomeron  may  indeed
enhance the signal to background  ratio.  However, care is needed with
this  argument.  Higgs  production  probes parton  distributions  at a
scale $Q^2 \sim M_H^2$, much larger than the typical  $Q^2$  scales of
diffractive  deep  inelastic  scattering at HERA.  Perturbative  DGLAP
evolution of the diffractive parton distributions to these high scales
gives rise to a mixing of the quark and gluon distributions such that,
for  example, a large  $g/q$  ratio at small  scales is washed  out at
higher  scales.  It is {\it a priori} not clear,  therefore,  that the
signal to background  ratio is enhanced in diffractive  events.  It is
precisely  this  question  that we wish to study here, using the three
models of pomeron structure presented in Ref.~\cite{Stirling96}.

In the  following  we  shall  present  numerical  results  for  single
diffractive   Higgs   production  at  the  future  CERN  LHC  collider
($\sqrt{s}=14$~TeV)  with the underlying  parton  distributions of the
pomeron as presented in \cite{Stirling96}.  These three pomeron models
are  obtained  from  fits  to  HERA  measurements  of the  diffractive
structure            function             $F_2^D(x,Q^2;x^{\rm\pom},t)$
\cite{H1lastyear,ZEUSlastyear},   and  are   described  in  detail  in
Ref.~\cite{Stirling96}.  In the present  context,  the most  important
distinguishing  feature of the models is the gluon distribution in the
pomeron, which differs significantly between them.  In summary:

\begin{description}
\item[Model~1:]{At  $Q_0=2$~GeV  the pomeron is entirely  composed  of
quarks.  Gluons are dynamically generated via DGLAP evolution.}
\item[Model~2:]{A  mix of quarks  and  gluons  at the  starting  scale
$Q_0$.}
\item[Model~3:]{A  predominantly  hard gluonic content at the starting
scale, the gluons in the pomeron  carry  large  fractional  momenta on
average.}
\end{description}

The   $Q^2$    evolution    of   the    gluon    distributions    $\xi
f_{g/{\rm\pom}}(\xi,Q^2)$,  with $\xi = x/x^{\rm\pom}$,  for the three
models is shown in Fig.~2.  The quark distributions are comparable for
all three models, being  constrained  by the HERA data, which explains
why the cross sections for diffractive $W^\pm$ and $Z$ production (via
$q  \bar  q  \to  W,Z$)  are  rather   similar  in  the  three  models
\cite{Stirling96},  in  contrast  to the Higgs  cross  sections  to be
presented below.

The dominant mechanism for Higgs production at the LHC is gluon--gluon
fusion    via    a    top    quark     loop,    see    for     example
Ref.~\cite{KMS}.\footnote{In  our  calculations  we also  include  the
direct $q\bar q \to H$  ($q=u,d,c,s,b)$  quark--fusion  processes, but
these are numerically much less important.}  The leading--order  cross
section is given by~\cite{Georgi78}
\beq \label{cross1}
\frac{d\sigma}{dy_H}   (pp{\rightarrow}  HX)  =  \sigma_0  \,  I\left(
\frac{m_t^2}{M_H^2}\right) \, f_{g/p}(x_1,Q^2) f_{g/p}(x_2,Q^2),
\eeq
with
\beq \label{cross0}
\sigma_0 = \frac{G_F\alpha_S^2(Q^2)}{32\sqrt{2}\pi} \frac{M_H^2}{s},
\eeq
for a Higgs boson of mass $M_H$ and rapidity  $y_H$.  The function $I$
in (\ref{cross1}) can be approximated by
\beq
I(x) \approx 1 + {1 \over 4x}, \qquad \mbox{for}\ x > 1 .
\eeq
The longitudinal momentum fractions of the gluons inside the colliding
protons   are   $x_{1,2}=(M_H/\sqrt{s})   e^{\pm   y_H}$.  The  single
diffractive  Higgs cross  section is obtained from  (\ref{cross1})  by
replacing one of the $f_{g/p}$ by the corresponding diffractive parton
distribution, i.e.
\beq \label{cross1SD}
\frac{d\sigma^{SD}}{dy_H}  (pp{\rightarrow}  HX) = \sigma_0\,  I\left(
\frac{m_t^2}{M_H^2}\right)\,         \left[         f^D_{g/p}(x_1,Q^2)
f_{g/p}(x_2,Q^2) + f_{g/p}(x_1,Q^2) f^D_{g/p}(x_2,Q^2) \right],
\eeq
where
\beq
f^D_{g/p}(x,\mu^2) = \int d \xpom d t\; {d f_{g/p}(x,\mu^2; \xpom , t)
\over d \xpom d t},
\label{eq:fDqp}
\eeq
with $d  f_{g/p}  / d \xpom d t$ given by  Eq.~(\ref{eq:fqp}).  In the
calculations which follow, the integration ranges are taken to be
\beq
0 \leq \xpom \leq 0.1, \qquad 0 \leq -t < \infty .
\label{eq:ranges}
\eeq
For the parton distributions $f_{i/p}(x,Q^2)$ in the proton we use the
MRS(A$'$)  set of partons  \cite{Martin95},  with QCD scale  parameter
$\Lambda_{\overline{\rm  MS}}^{N_f=4} = 231$~MeV, which corresponds to
$\alpha_S(M_Z^2)  = 0.113$.  At the level of  accuracy to which we are
working, all modern parton distribution sets give essentially the same
results.  The  renormalization/factorization  scale  is  chosen  to be
$Q^2=M^2_H$.  We use  leading--order  expressions  for the  signal and
background  cross sections, since our primary  interest is in the {\it
ratio} of  diffractive  to total cross  sections,  which should not be
significantly  affected  by  higher--order  corrections  to the  basic
subprocesses.  In any case, the diffractive  parton  distribution fits
to the deep inelastic data do not yet require NLO corrections.

The cleanest  decay channel for searching  for the  intermediate  mass
Higgs   boson   at   the   LHC   is   $H   \to   \gamma\gamma$,   with
$\rm{Br}(H\rightarrow   \gamma\gamma)  \sim  3\times   10^{-4}-3\times
10^{-3}$   for   $50$~GeV$\leq   M_H\leq   150$~GeV   \cite{KMS}.  The
irreducible  background  comes  from the  $O(\alpha^2)$  $q\bar  q \to
\gamma\gamma$ and the  $O(\alpha^2\alpha_s^2)$  $gg \to  \gamma\gamma$
\cite{Combridge80}  subprocesses.  Note that these provide {\it lower}
bounds to the background  cross sections, since reducible  backgrounds
from e.g.  $q g \to \gamma q (q \to \gamma,  \pi^0,  \ldots)$ can also
be important in practice, see for example  Ref.~\cite{Stirling89}.  In
what follows we will ignore these additional  contributions,  assuming
that they can be  suppressed  by photon  isolation  cuts.  For  larger
Higgs masses, i.e.  for $M_H > 2 M_Z$, the important  decay channel is
$ H \to ZZ\to 4 l ^\pm$, with  $\rm{Br}(H\rightarrow  ZZ) \approx 0.3$
\cite{KMS}.  In this range, the  dominant  irreducible  background  is
from $q \bar q \to ZZ$.

In Fig.~3a we show the total  (\ref{cross1})  and  single  diffractive
(\ref{cross1SD}) Higgs cross sections, the latter calculated using the
three sets of pomeron parton distributions of  Ref.~\cite{Stirling96}.
As  expected,  Model~3  with  the  largest  gluon  gives  the  largest
diffractive  cross  section.  Model~1  has  no  gluons  at  all at the
starting scale $Q_0=2$~GeV;  gluons are dynamically  created via DGLAP
evolution at higher  values of $Q$.  However,  the gluon  distribution
remains  quite  small  compared to Models~2  and 3.  Taking the models
together,  we see that  between  approximately  2\% and  15\% of Higgs
events are expected to be singly  diffractive.\footnote{Recall that we
impose a cut $\xpom \leq 0.1$ when  calculating the diffractive  cross
sections.}  Our  results for the single  diffractive  and total  Higgs
cross    sections   are    consistent    with   those    obtained   in
Ref.~\cite{GRAUDENZ} using similar models.

Fig.~3b shows the $\gamma\gamma$  background for the lower part of the
mass range, with $M_H$ now  replaced by the  $\gamma\gamma$  invariant
mass  $M_{\gamma\gamma}$.  Note that in {\it both}  Figs.~3a and 3b we
impose a cut of $\vert y_\gamma\vert \leq 2$ to approximately  account
for the  experimental  acceptance.  As the inset in Fig.~3b shows, the
gluon--gluon    fusion    process    dominates    for    very    small
$M_{\gamma\gamma}$  where small parton momentum  fractions are probed.
The $q\bar{q}$ subprocess dominates at large  $M_{\gamma\gamma}$.  The
corresponding  single diffractive cross sections are again largest for
the  gluon--richer  pomeron  models, in particular  Model~3.  However,
even the gluon--poor  Model~1 becomes comparable to Model~2 due to the
increasing  $q\bar{q}$  contribution  to the  cross  section  at large
$M_{\gamma\gamma}$.

The $ZZ$  backgrounds,  relevant for higher Higgs masses, are shown in
Fig.~4b.  We see that in contrast to the $\gamma\gamma$ backgrounds of
Fig.~3b, all three pomeron models give  comparable  diffractive  cross
sections  over  the  entire  $M_{ZZ}$   range.  This  is  because  the
diffractive quark  distributions are constrained to be the same by the
HERA $F_2^D$ data.

Before  discussing  the single  diffractive  ratios of the  signal and
background  processes, it is  interesting  to study in more detail the
kinematics of diffractive Higgs  production, in particular the typical
values of the various momentum  fractions in the calculation.  Thus in
Fig.~5 we show the average gluon momentum fraction $\langle  x\rangle$
inside the pomeron, the momentum fraction $x^{\rm\pom}$ of the pomeron
and   the    average    value    of   the    variable    $\xi$    with
$\langle\xi\rangle=\langle  x/x^{\rm\pom}\rangle$,  as a  function  of
$M_H$.  The  calculation  of these  quantities  allows the Higgs cross
sections  in  the  different  models  to  be  related  to  the  parton
distributions  of  Fig.~2.  The  gluon  momentum  fraction  shows  the
typical  $\langle  x\rangle  \propto  M_H/\sqrt{s}$   behaviour  which
follows  from the input  $x_{1,2}=(M_H/\sqrt{s})e^{\pm  y_H}$  for the
momentum  fractions  of the gluons in $gg \to H$,  Eq.~(\ref{cross1}).
The  fractional  pomeron  momentum  is of  course  constrained  to  be
$x^{\rm\pom}\leq  0.1$ and it stays  very  close to this  upper  limit
throughout  the complete range of $M_H$.  It exhibits an almost linear
but  very  weak  $M_H$  dependence  for  $M_H\geq  100$~GeV:  $\langle
x^{\rm\pom}\rangle\sim  3.2\times  10^{-5}M_H$.  The relevant variable
for  comparison  with the  parton  distributions  in Fig.~2  is $\xi =
x/x^{\rm\pom}$.  For  light  Higgs  masses  the  values  for  $\langle
\xi\rangle$ are small,  $(\langle\xi\rangle < 0.1$ for $M_H<100$~GeV).
In this region of $Q = M_H$ Models~2 and 3 (cf.  Figs.~2b and 2c) have
approximately the same gluon content, which explains the similarity of
the  corresponding  diffractive cross sections in Fig.~3a.  For higher
values of $M_H$, the difference  between  Model~2 and Model~3  becomes
more  apparent:  the gluon  distribution  in Model~3  remains  roughly
constant,  while that of Model~2  decreases for higher values of $M_H$
and $\xi$.  This explains the  differences  between  Models~2 and 3 in
Figs.~3a and 4a.  We assume that the kinematics  illustrated in Fig.~5
for the Higgs cross sections are also valid for the $\gamma\gamma$ and
$ZZ$ backgrounds at the equivalent invariant mass.

Finally    we    present     the     single     diffractive     ratios
$R^{SD}=\sigma^{SD}/\sigma$  for the signal  ($pp\rightarrow H+X$) and
background     contributions      ($pp\rightarrow\gamma\gamma     +X$,
$pp\rightarrow ZZ+X$) to see whether the signal to background ratio is
indeed enhanced by the gluon--rich  pomeron.  Fig.~6a shows the ratios
for the  Higgs  mass  range  $M_H\leq  200$~GeV.  For the  gluon--rich
Models~2 or 3, there is indeed a slight  enhancement  of $R^{SD}$  for
the signal  compared to the  background,  for example in Model~3 for a
Higgs mass of  $M_H=100$~GeV  we find  $R^{SD}_H\sim14\%$  compared to
$R^{SD}_{\gamma\gamma}\sim11\%$.  The  enhancement  persists  over the
whole  Higgs  mass  range.  For the  gluon---poor  Model~1,  where the
gluons are  dynamically  produced by DGLAP  evolution,  the background
ratio is larger  than the signal  ratio for  $M_H>70$~GeV.  This small
enhancement  has to be contrasted with the (at least) factor of 5 loss
in the overall production rate.

The  situation  becomes  even more  dramatic if we go to higher  Higgs
masses  ($200$~GeV$\leq  M_H\leq  1000$~GeV)  as shown in Fig.~6b.  In
this case the important  background to Higgs production is direct $ZZ$
pair production via quark--antiquark annihilation, as discussed above.
As expected, in Model~1 the background  ratio exceeds the signal ratio
by  a  large  factor   ($\approx  6$  for   $M_H=200$~GeV).  Even  the
gluon--richer  Model~2  yields a higher  background  contribution  for
$M_H<350$~GeV.  Only at higher  masses  (i.e.  evolution  scales)  are
enough  additional  gluons  produced to enhance  the signal.  Only the
very  gluon--rich  Model~3,  with enough  gluons  even at low  scales,
allows for a dominant signal ratio throughout the entire mass range.

In  conclusion,  we have  calculated  single  diffractive  Higgs cross
sections for the LHC using diffractive parton  distributions  based on
quark and gluon  constituents  of the pomeron,  fitted to HERA $F_2^D$
data.  In particular, we have considered  three models which differ in
the  relative  amounts  of  quarks  and  gluons.  If  the  pomeron  is
gluon--rich,  then between 5\% and 15\%  (depending on the Higgs mass)
of Higgs events should have a single diffractive  structure.  Assuming
the overall validity of this `universal pomeron structure' model, more
precise  measurements  of  $F_2^D$  at HERA will  allow more  accurate
predictions.  However we have also shown that there is no  significant
enhancement  of the  signal to  background  ratio in such  diffractive
events.  DGLAP  evolution  to high  scales $Q \sim M_H$  automatically
generates a mixture of diffractive quark and gluon  distributions, and
so the  background  processes $q \bar q, gg \to  \gamma\gamma$  and $q
\bar  q\to ZZ$  also  have a large  diffractive  component.  It is not
clear,  therefore,  that there is any advantage in searching for Higgs
bosons at the LHC in events with rapidity gaps.

\section*{Acknowledgements}

\noindent We thank Dirk  Graudenz  for useful  discussions.  This work
was also  supported  in part by the EU  Programme  Human  Capital  and
Mobility,  Network  `Physics  at  High  Energy  Colliders',   contract
CHRX-CT93-0357  (DG 12 COMA).  MH  gratefully  acknowledges  financial
support in the form of a DAAD-Doktorandenstipendium (HSP III).

\goodbreak


\begin{figure}[t]             
\unitlength1cm            
\begin{center}
\begin{picture}(10,12)            
\makebox[9.5cm]{\epsfxsize=9cm
\epsfysize=12cm 
\epsffile{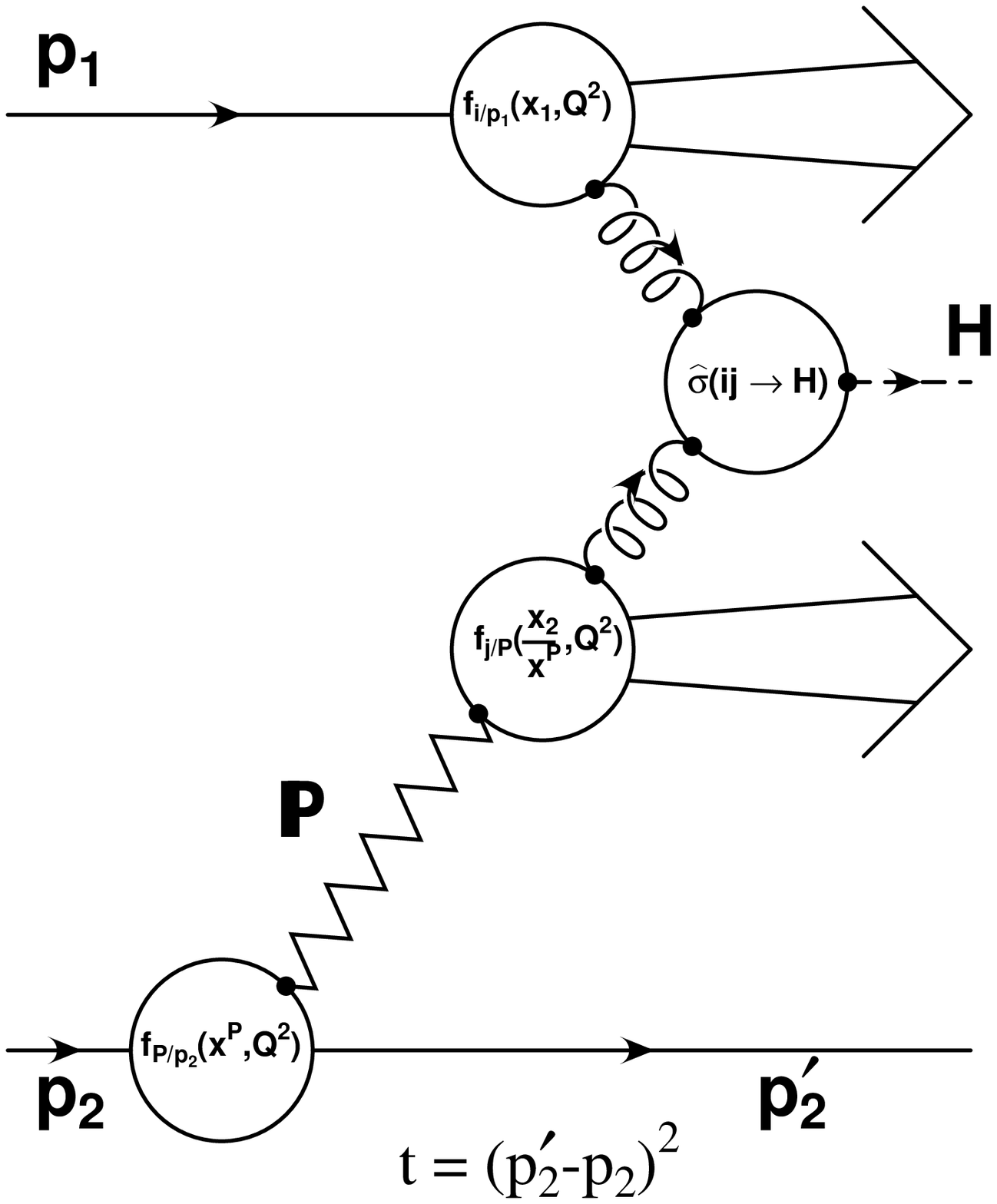} } 
\end{picture}
\end{center}

\caption[]{Kinematics  of a single  diffractive  hard scattering event
for  two  colliding  protons,  with  $p_2$  being   quasi--elastically
scattered  by  emitting  a  pomeron   {\rm\bigpom}   which   undergoes
interaction with $p_1$ to create, for example, a Higgs boson $H$.}

\end{figure}


\begin{figure}[t]             
\unitlength1cm            
\begin{center}
\begin{picture}(13,16)            
\makebox[12.5cm]{\epsfxsize=12cm
\epsfysize=16cm 
\epsffile{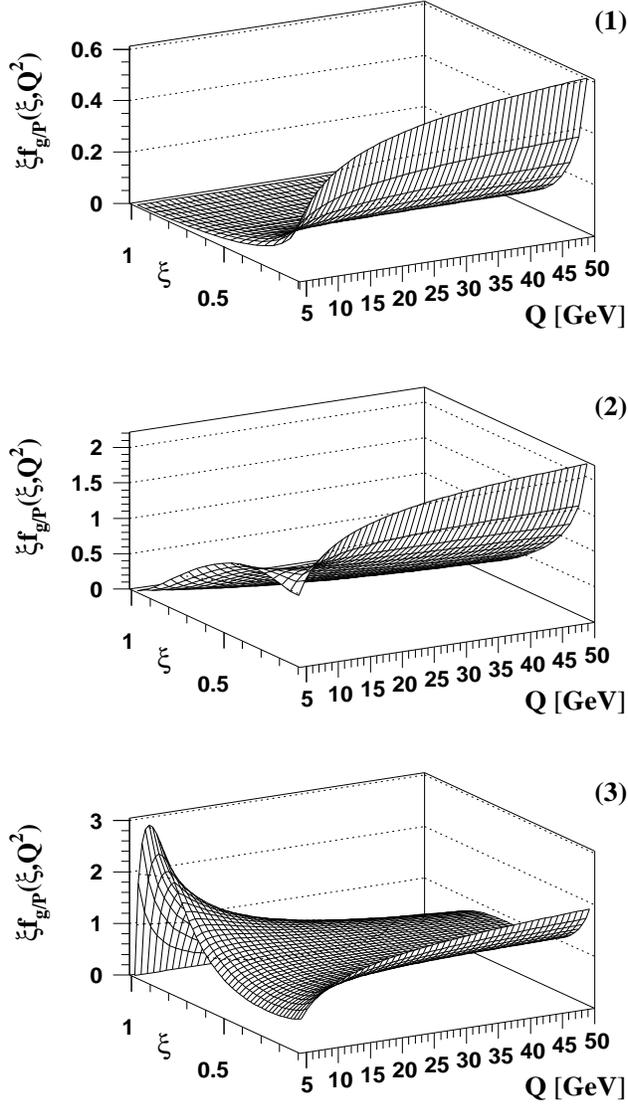} } 
\end{picture}
\end{center}

\caption[]{The   $Q^2$  evolution  of  the  gluon  distributions  $\xi
f_{g/\pom}(\xi,Q^2)$  in the three different pomeron  structure models
of Ref.~\cite{Stirling96}.}

\end{figure}


\begin{figure}[t]             
\unitlength1cm            
\begin{center}
\begin{picture}(13,16)            
\makebox[12.5cm]{\epsfxsize=12cm
\epsfysize=16cm 
\epsffile{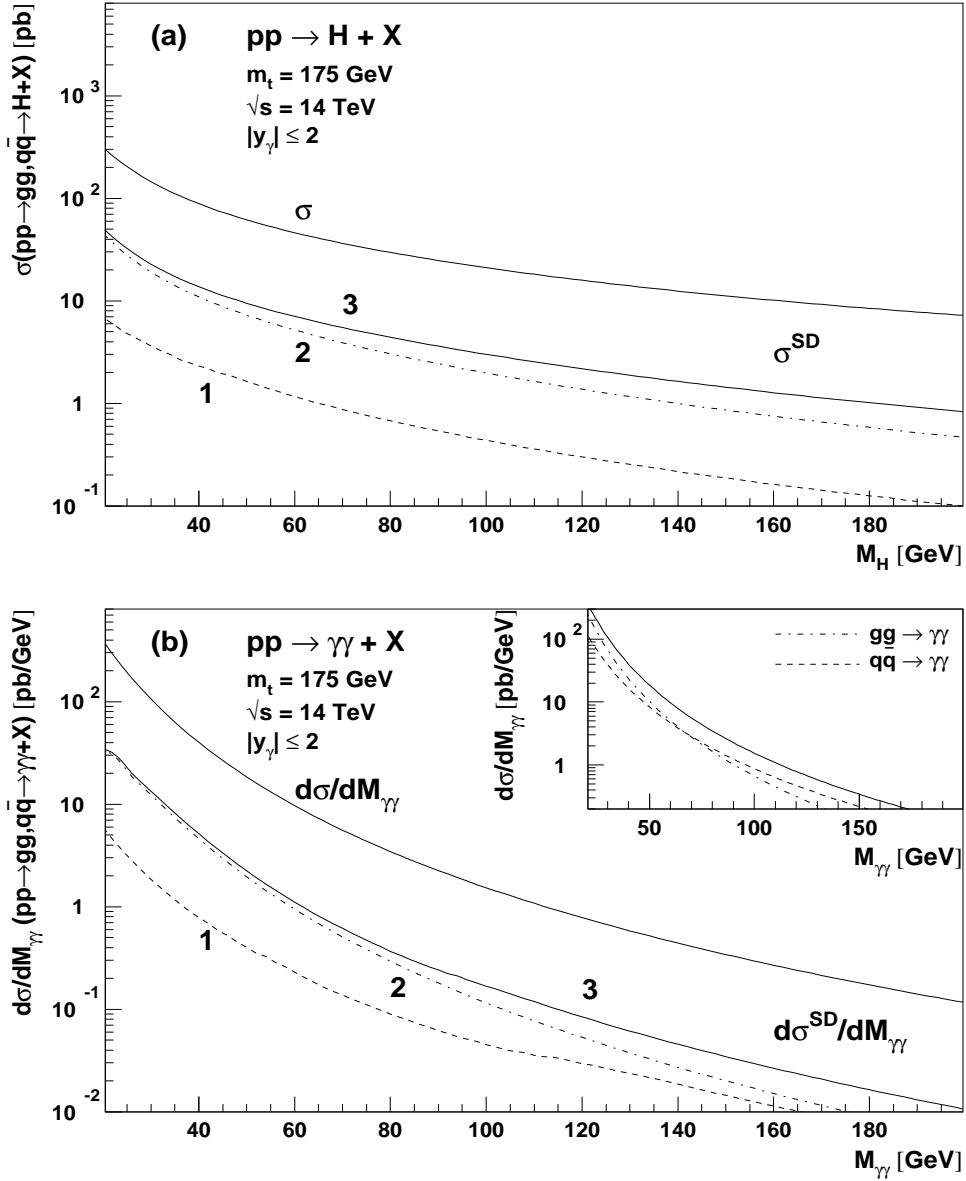} } 
\end{picture}
\end{center}

\caption[]{The total and the single diffractive cross sections for (a)
Higgs  production  as a  function  of the  Higgs  mass  $M_H$  and (b)
$\gamma\gamma$   production   as   a   function   of   the   invariant
photon--photon mass $M_{\gamma\gamma}$ for the three different pomeron
models of Ref.~\cite{Stirling96}.  For both signal (assuming the decay
$H\rightarrow\gamma\gamma$)  and background the photons are restricted
to the central region by a cut $|y_{\gamma}|\leq 2$.  The inset in (b)
shows the leading  order cross section and the relative  contributions
from gluon--gluon fusion and quark--antiquark annihilation.}

\end{figure}


\begin{figure}[t]             
\unitlength1cm            
\begin{center}
\begin{picture}(13,16)            
\makebox[12.5cm]{\epsfxsize=12cm
\epsfysize=16cm 
\epsffile{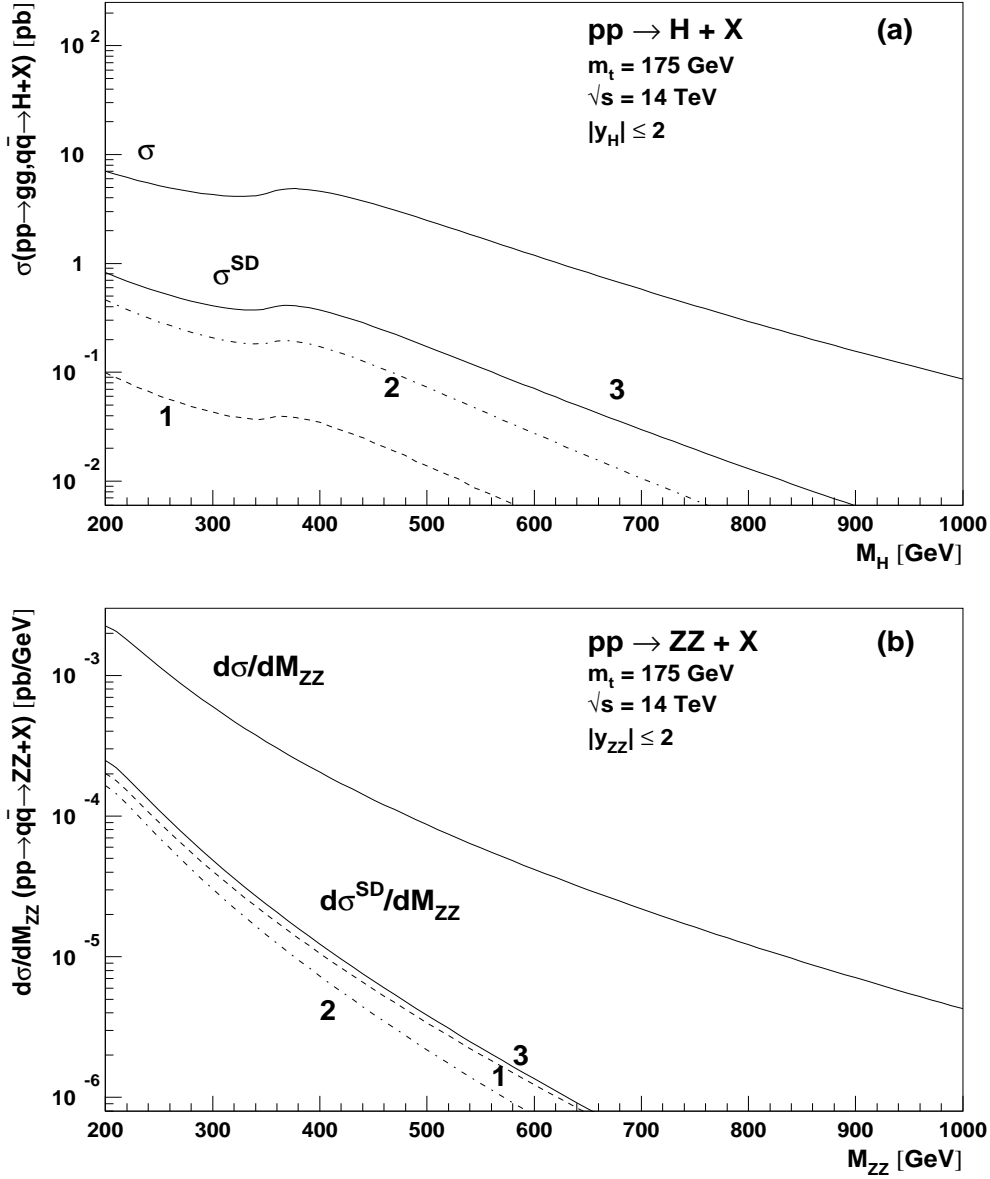} } 
\end{picture}
\end{center}

\caption[]{The total and the single diffractive cross sections for (a)
Higgs  production  as a function  of the Higgs mass $M_H$ and (b) $ZZ$
production as a function of the invariant  $ZZ$ mass  $M_{ZZ}$ for the
three different  pomeron models of  Ref.~\cite{Stirling96}.  The Higgs
and the  $ZZ$  pair  are  restricted  to the  central  region  by cuts
$|y_{ZZ}|,\/ |y_H| \leq 2$.}

\end{figure}


\begin{figure}[t]             
\unitlength1cm            
\begin{center}
\begin{picture}(13,8)            
\makebox[12.5cm]{\epsfxsize=12cm
\epsfysize=8cm 
\epsffile{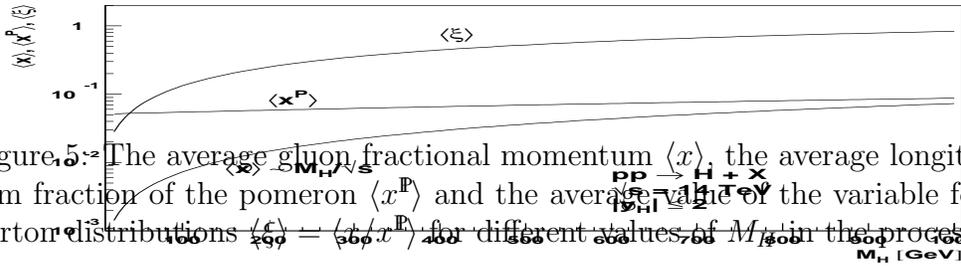} } 
\end{picture}
\end{center}

\vskip-5cm

\caption[]{The  average gluon fractional momentum $\langle  x\rangle$,
the average  longitudinal  momentum  fraction of the pomeron  $\langle
x^{\rm\pom}\rangle$  and the  average  value of the  variable  for the
pomeron   parton   distributions   $\langle   \xi\rangle   =   \langle
x/x^{\rm\pom}\rangle$ for different values of $M_H$ in the process $pp
\rightarrow H+X$.}

\end{figure}


\begin{figure}[t]             
\unitlength1cm            
\begin{center}
\begin{picture}(13,16)            
\makebox[12.5cm]{\epsfxsize=12cm
\epsfysize=16cm 
\epsffile{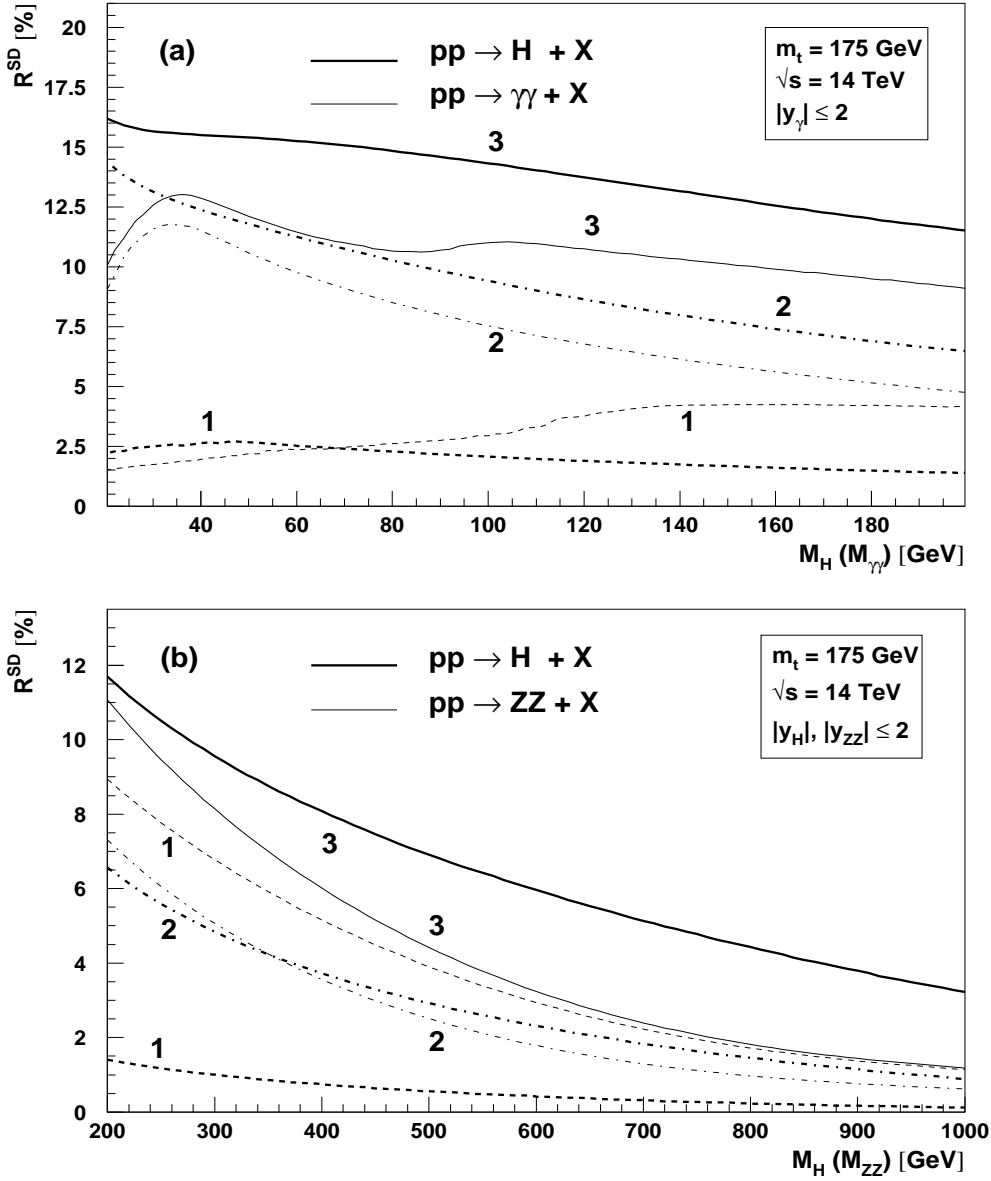} } 
\end{picture}
\end{center}

\caption[]{The  single diffractive ratios  $R^{SD}=\sigma^{SD}/\sigma$
for $pp\rightarrow H+X$ (solid lines) and the background contributions
(a)  $pp\rightarrow  \gamma\gamma + X$ and (b)  $pp\rightarrow ZZ + X$
(dashed lines) for the three different  pomeron  models.  The absolute
values of the cross sections $\sigma$ and $\sigma^{SD}$  are presented
in Figs.~3 and 4.}

\end{figure}

\end{document}